\documentclass{emulateapj}

\begin{document}

\title{A Robust Determination of the size of quasar accretion disks\\using gravitational 
microlensing}

\author{J. Jim\'enez-Vicente}
\affil{Departamento de F\'{\i}sica Te\'orica y del Cosmos. Universidad de Granada, Campus de Fuentenueva, 18071, Granada, Spain \\
Instituto Carlos I de F\'{\i}sica Te\'orica y Computacional. Universidad de Granada, 18071, Granada, Spain}
\author{E. Mediavilla}
\affil{Instituto de Astrof\'{\i}sica de Canarias, V\'{\i}a L\'actea S/N, La Laguna 38200, Tenerife, Spain \\
Departamento de Astrof\'{\i}sica, Universidad de la Laguna, La Laguna 38200, Tenerife, Spain}
\author{J. A. Mu\~noz}
\affil{Departamento de Astronom\'{\i}a y Astrof\'{\i}sica, Universidad de Valencia, 46100 Burjassot, Valencia, Spain}
\author{C. S. Kochanek}
\affil{Department of Astronomy, The Ohio State University, 140 West 18th Avenue, Columbus, OH 43210, USA\\
Center for Cosmology and Astroparticle Physics, The Ohio State University, 191 West Woodruff Avenue, Columbus, OH 43210, USA}

\begin{abstract}

Using microlensing measurements from a 
sample of 27 image-pairs of 19 lensed quasars we determine a maximum
likelihood estimate for the accretion disk size of an {{\em}average} quasar of $r_s=4.0^{+2.4}_{-3.1} $ light days at rest frame ${\langle}\lambda{\rangle}=1736$\AA\ for microlenses with a mean mass of ${\langle}M{\rangle}=0.3M_\odot$. 
This value, in good agreement with previous results from
smaller samples, is roughly a factor of 5 greater than
the predictions of the standard thin disk model.
The individual size estimates for the 19 quasars in our sample 
are also in excellent agreement with the results of the joint maximum likelihood analysis.

\end{abstract}

\keywords{accretion, accretion disks --- gravitational lensing: micro --- quasars: general}

\section{Introduction}

The thin disk (Shakura \& Sunyaev 1973) is the standard model to describe the inner regions of quasars. This
model predicts typical sizes for the accretion disks of luminous quasars of $\sim$$10^{15}\,\rm{cm}$ (${\sim}0.4$ light
days) in the (observer frame) optical bands. However, recent measurements based on quasar microlensing  are challenging this prediction. 
Based on a combined optical and X-ray study of
microlensing in 10 quadruply lensed quasars, Pooley et al. (2007) found that the source of optical light is
larger than expected from basic thin accretion disk models by factors of 3 to 30. Studying microlensing
variability observed for 11 gravitationally lensed quasars Morgan et al. (2010) also found that the
microlensing estimates of the disk size are larger (by a factor ${\sim}4$) than would be expected from thin disk
theory. Other monitoring based studies 
focused on individual objects like QSO2237+0305 (Kochanek
2004, Anguita et al. 2008, Eigenbrod et al. 2008, Mosquera et al. 2009,  Agol et al. 2009, Poindexter \& Kochanek 2010),
PG1115+080 (Morgan et al. 2008), RXJ1131$-$1231 (Dai et al. 2010), HE1104$-$1805 (Poindexter et al. 2008) or HE0435$-$1223 (Blackburne et al. 2011b) generally support these conclussions. 

From single-epoch optical/IR and X-Ray measurements,
Blackburne et al. (2011a) used the wavelength dependence of microlensing (chromaticity) to study the structure
of accretion disks in 12 4-image lensed quasars, finding disk sizes larger than predicted by
nearly an order of magnitude. More detailed studies of individual objects based in microlensing chromaticity
(Eigenbrod et al. 2008, Poindexter et al. 2008, Mosquera et al. 2011, Mediavilla et al. 2011a, Mu\~noz et al. 2011) also found that the size estimated from
microlensing is substantially larger than the values predicted by the thin disk theory (by factors of about 10 in
HE0435$-$1223, 5 in SBS0909+532, and 4 in HE1104$-$1805). Upper limits
on the sizes of MG0414+0534 and SDSSJ0924+0219 have also been set by Bate et al. (2008) and Floyd et al. (2009), respectively. 

Here, we
will generalize these results by extending the study to the sample of 20 lensed quasars of 
Mediavilla et al. (2009, hereafter MED09).
Our data consists of microlensing amplitude measurements for 29 image-pairs from 20 lensed quasars
based on a comparison between the flux ratios in the continuum and an adjacent emission line. The emission line flux ratio provides an unmicrolensed baseline that also removes the effects of extinction (see e.g. Falco et al. 1999, Mu\~noz et al. 2004) and 
lens
mo\-dels in the  determination of the magnification anomalies. 
As noted in previous works (Kochanek 2004,
Eigenbrod et al. 2008, Blackburne et al. 2011a), studies favouring the selection of objects with noticeable magnification
anomalies (or epochs with high microlensing activity) may lead to a bias towards smaller quasar size determinations.
The fact that one
object does not show microlensing magnification anomalies may be due to the fact that it is either too big or that it lies in a region without significant magnification
fluctuations. Thus, non-detection of microlensing also puts constraints on the size
of the sources and should be taken into account.
The MED09 sample is also unaffected by this bias (indeed the histogram of estimated microlensing magnifications in this sample peaks at $\Delta m=0$).
However, the data from MED09 do not include, for most image-pairs, enough
measurements at different wavelengths to address chromaticity. For this reason we will start discussing the
impact of chromaticity on the determination of microlensing based sizes (\S 2) to show that size estimates are
dominated by microlensing amplitude and are relatively independent of chromaticity. In \S3 we will use the microlensing data from MED09 to calculate a maximum likelihood estimate of the accretion disk
size. 
The estimate will be for the typical size of the accretion disk of an {\em average} quasar rather than that of any particular source. 
We also make individual size estimates for the 19 objects using a complementary
Bayesian approach, finding excellent agreement with the likelihood analysis. 
The main conclusions are presented in \S 4. 

\section{Size determination from single wavelength microlensing measurements}

We will base our analysis on the microlensing measurements inferred by MED09 from
optical spectroscopy available in the literature (29 quasar image pairs seen through 20 lens galaxies).  
The microlensing 
magnification between two images 1 and 2 is
calculated as $\Delta m=(m_2-m_1)_{micro}=(m_2-m_1)_{cont}-(m_2-m_1)_{line}$, where  $(m_2-m_1)_{line}$ is the flux ratio in an emission line and
$(m_2-m_1)_{cont}$ is the flux ratio of the continuum adjacent to the emission line.  
With this method the microlensing magnification is isolated from extinction and from the mean lensing magnification, as these effects affect the line and the adjacent continuum equally. 
The consistency of this procedure has been confirmed with mid-IR data (MED09).  

The histogram of observed microlensing
magnifications in MED09 showed that 93\% of the quasar image pairs have microlensing 
$|\Delta m|\le{0.8}\rm\,{mag}$. In this section, we are
going to use this statistical constraint to estimate quasar accretion disk sizes from single wavelength microlensing measurements.

We will apply a statistical approach si\-mi\-lar to that used
by \cite{Mediavillaetal11} 
to a representative lens system. 
Therefore, we take for the lens a standard 
SIS with values $\kappa_1=\gamma_1=0.45$ 
and $\kappa_2=\gamma_2=0.55$ for the convergence and shear at the positions of the two images. 
These values are typical of the lenses in the MED09 (cf. their Table 4). For the redshifts of lens and source we have taken the values 
$z_l=0.57$ and $z_s=1.76$, which are the average values for the sample.
\begin{figure*}[t]
\epsscale{1.1}
\plotone{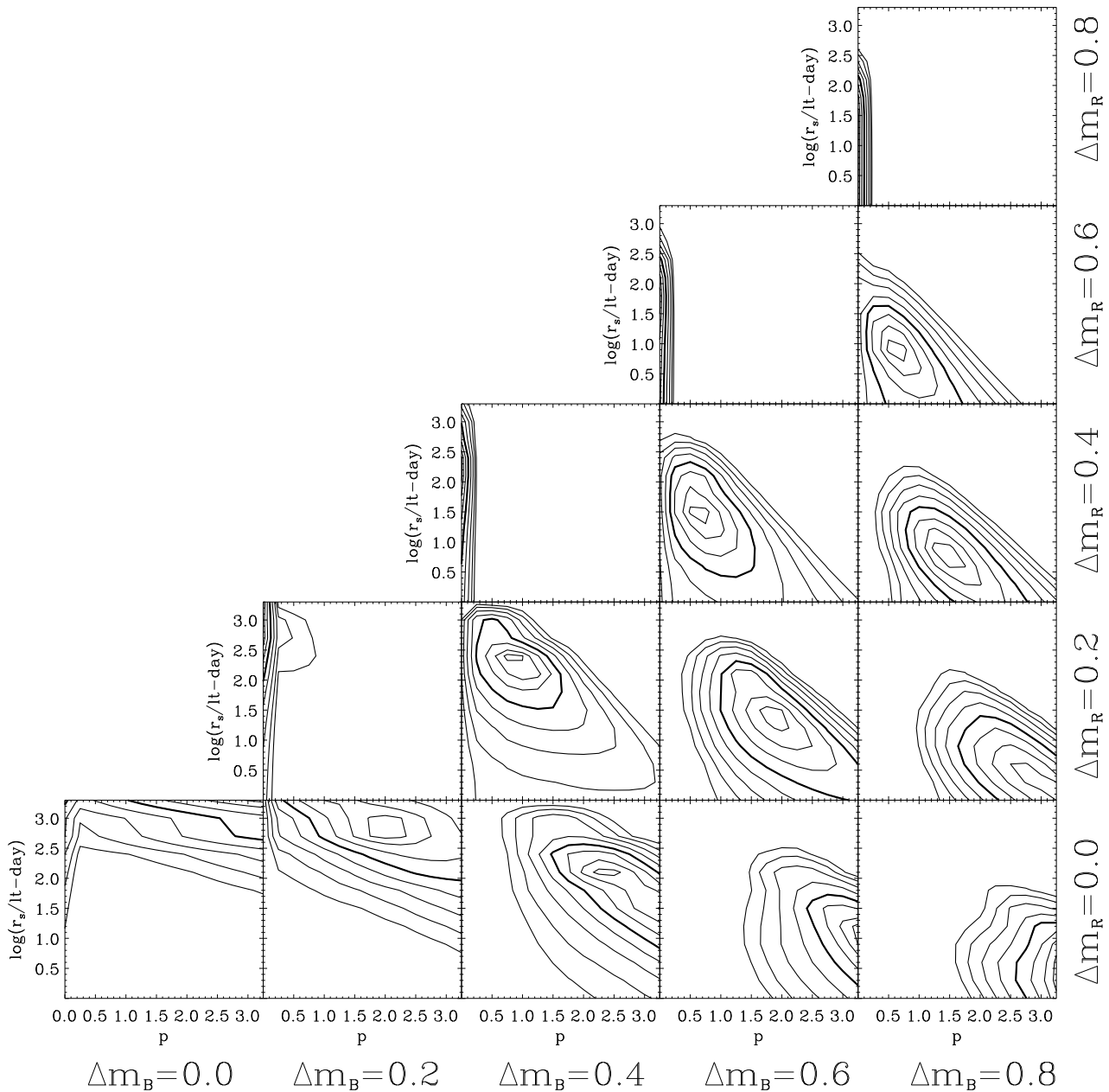}
\caption{Grid of two dimensional probability distributions $p_{r^i_s, p^j}(\Delta m_B^{obs},{\Delta}m_R^{obs})$ for the 15 cases with ${\Delta}m_B>0$ and $\alpha=0.05$. Each column corresponds to the distribution for a given value of ${\Delta}m_B$ indicated at the top of the column. Each row corresponds to  the distribution for a given value of ${\Delta}m_R$ indicated at the right. For each distribution the abcissae represent $p$ from 0.0 to 3.25 and the ordinates represent ${\log}\,r_s$ from 0.0 to 3.3 ($r_s$ in light days). The contour levels are drawn at intervals of $0.25\sigma$ (the contour at $n\sigma$ is drawn at $\exp(-n^2\sigma^2/2)$ from the peak of the distribution). The contour at $1\sigma$ is thicker than the rest.\label{fig1}}
\end{figure*}

Using these parameters for the macro lens, we ge\-ne\-ra\-te microlensing magnifications
maps at the positions of images 1 and 2 using the inverse polygon mapping algorithm described in 
\cite{MediavillaetalIPM1} and \cite{MediavillaetalIPM2}.
The calculated maps are $2000{\times}2000$ pixels in size, with a pixel size of 0.5 light days, 
or equivalently $1.295{\times}10^{15} $cm. In the image plane, the covered area is the source plane 
area increased by factors $1.5(1-\kappa\pm\gamma)^{-1}$.
For the fraction of projected mass density in stars, $\alpha$, we use 
current estimates (Pooley et al. 2009, MED09, Mosquera et al. 2011) and
consider only the cases with
$\alpha=0.05$ and $\alpha=0.1$. We used microlenses of $1M\sun$.

In order to describe the structure of the source,
we model the accretion disk with a Gaussian profile 
$I(R)\propto\exp(-R^2/2r_s^2)$, where the characteristic size $r_s$ is a wavelength dependent 
parameter (related to the half-light radius as $R_{1/2}=1.18r_s$). The wavelength dependence of the disk size is parametrized by a power law with
exponent $p$ such that $r_s(\lambda)\propto\lambda^p$.
We smooth the magnification maps with Gaussians scaled to
two wavelengths 
$\lambda_B$ and $\lambda_R$ representing characteristic blue and red wavelengths. We choose 
$\lambda_R/\lambda_B=2.3$, the ratio between $\mathrm{Ly}\alpha$ and Mg II, which is roughly the wavelength range covered by observations
in the optical band. 

We calculate the probability of observing microlensing magnifications ${\Delta}m=\mu_2-\mu_1$ (where $\mu_1$ and $\mu_2$ are the magnifications at the positions of image 1 and 2 respectively) at wavelengths $\lambda_B$ and $\lambda_R$, namely, ${\Delta}m_B$ and ${\Delta}m_R$, for a grid in the size at $\lambda_B$, $r_s(\lambda_B)$, and the exponent $p$. We use a lo\-ga\-rith\-mic grid in $r_s$  such that ${\log}\,r^i_s=0.3{\times}i$ for $i=0\cdots 11$ ($r^i_s$ in light days) and a linear grid in $p$ such that $p^j=0.25{\times}j$ for $j=0\cdots 13$. 
We restrict ourselves
to the case $|{\Delta}m_B|>|{\Delta}m_R|$, as we assume that the size of the source increases
with wavelength, and therefore the largest magnifications are expected for the smallest (i.e. bluest) 
sources\footnote{Inversions are possible but rare (see Poindexter et al. 2008)}.

The pro\-ba\-bi\-li\-ty of observing ${\Delta}m_B^{obs}$ and ${\Delta}m_R^{obs}$  in a model with parameters $r^i_s$ and $p^j$ is 
given by
\begin{equation}
p_{r^i_s, p^j}(\Delta m_B^{obs},{\Delta}m_R^{obs})\propto{\int}d{\Delta}m_B{\int}d{\Delta}m_RN_{ij}e^{-\frac{1}{2}\chi^2},
\end{equation}
where
\begin{equation}
\chi^2=\frac{({\Delta}m_B-{\Delta}m_B^{obs})^2}{\sigma^2_{{\Delta}m_B^{obs}}}+\frac{({\Delta}m_R-{\Delta}m_R^{obs})^2}{\sigma^2_{{\Delta}m_R^{obs}}}
\end{equation}
Here, $N_{ij}$ is the number of trials with ${\Delta}m_B$ and ${\Delta}m_R$ for the case 
with parameters $r^i_s$ and $p^j$, and $\sigma_{{\Delta}m_B^{obs}}$ and  $\sigma_{{\Delta}m_R^{obs}}$ are
the uncertainties in the observed mi\-cro\-len\-sing magnifications, for which we have taken a 
typical value of  $\sigma_{{\Delta}m_B^{obs}}=\sigma_{{\Delta}m_R^{obs}}=$0.05 mags.

The calculations are performed for a grid of pairs ${\Delta}m_B, {\Delta}m_R$ with ${\Delta}m_B=0.2{\times}k$ for $k=-4\cdots 4$ and ${\Delta}m_R=0.2{\times}m$ for $m=0\cdots k$ (so we always have $|m|<|k|$ and  $|{\Delta}m_B|>|{\Delta}m_R|$) for $\alpha=0.05$ and $\alpha=0.1$. This range of values for 
the strength of mi\-cro\-len\-sing covers the vast majority of observed cases as shown in the histogram of observed microlensing strengths 
in the sample of MED09 (their Fig. 1). 
Thus, the pro\-ba\-bi\-li\-ty distribution $p_{r^i_s, p^j}(\Delta m_B^{obs},{\Delta}m_R^{obs})$ is
calculated for $14{\times}12{\times}29{\times}2=9744$ cases. For each one of those cases (i.e. for every quadruplet $(p, r_s,\Delta m_B^{obs},{\Delta}m_R^{obs})$), the probability is 
calculated using $10^8$ trials.

The results for the case  $\alpha=0.05$ are shown in Figure \ref{fig1} for ${\Delta}m_B>0$.
We have shown only the right side of the full figure, as the cases with ${\Delta}m_B<0$ are 
very similar to those with the same value of $|{\Delta}m_B|$ so that the scenario is fairly 
symmetric with 
respect to ${\Delta}m_B=0$. For the case $\alpha=0.1$ the results are very similar but the peaks
in the probability distributions are displaced towards slightly higher values of $r_s$. 

With the exception of the cases in which ${\Delta}m_B={\Delta}m_R$, the distributions show a clear
covariance in the sense that larger sizes imply lower values of the exponent $p$.
We also see that larger values of ${\Delta}m_B$ favour smaller values for the source size $r_s$. 

The most important result from this figure is that the probability distribution with respect to the size $r_s$ is dominated by the 
strength of microlensing (${\Delta}m_B$) with very little dependence on the amount of chromaticity. Conversely, the
value of the chromaticity (${\Delta}m_B-{\Delta}m_R$) is the main factor in determining 
the exponent $p$. 
Indeed, the most probable
values of $r_s$ (values within 1$\sigma$ around the maximum of the distributions in Figure \ref{fig1}) remain roughly constant along each column in Figure \ref{fig1} as we move through
the different rows. That means that even if we lack information on chromaticity, we can still obtain 
valuable  information on the size of the sources from measurements of microlensing
at a single wavelength. 
Another interesting and robust result from Figure \ref{fig1} is that, 
with the exception of the case with no detected microlensing (${\Delta}m_B=0$), the maximums of the distributions are all located in a fairly restricted range in $r_s$. Therefore, 
microlensing magnifications in the range $0<|{\Delta}m_B|\leq 0.8$ 
will provide sizes in a rather restricted range between ${\sim}1\sqrt{M/M_\odot}$ and ${\sim}16\sqrt{M/M_\odot}$ light days irrespective of chromaticity, with larger magnifications favouring smaller sizes.

\section{Discussion}

Based on the results of the previous section, it is possible to constrain 
the size of the sources from measurements of the microlensing strength at a single
wavelength, even if we
have no information on the amount of chromaticity present.
As noted above (see also Kochanek 2004,
Eigenbrod et al. 2008, Blackburne et al. 2011a), each single 
microlensing measurement is affected by a de\-ge\-ne\-ra\-cy between
the range of possible magnifications and source size effects.
Statistics of large samples of lensed quasars (like MED09) can be used to minimize this
uncertainty. In particular, we are going to use a maximum likelihood method to estimate an average size from the product
of the individual likelihood functions for each image pair. 
This method includes the cases with little or no
microlensing that by themselves would give rise only to lower limits on the size on an equal footing with those 
showing significant microlensing.

The microlensing magnifications in this sample  have been calculated using the
continuum to line flux ratios of different
lines for different objects. As size is related to wavelength, we should ideally use objects with 
magnifications obtained
from lines at similar wavelengths. On the other hand, to have good statistics, 
we should also try to keep the sample as large as 
possible.
With these restrictions in mind, 
we have chosen a compromise in which we
used all objects with magnifications measured in the wavelength range between Ly$\alpha$ (1216\AA) to Mg II (2798\AA). With this choice, the average rest wavelength is in the relatively narrow range of $\langle\lambda\rangle=1736\pm373$\AA, while we still keep 27 image pairs from 19 lensed quasars. The dispersion introduced by the effect of wavelength in the size estimate will be taken into account later.
We then use the measured microlensing magnifications in these 27 pairs as observed 
${\Delta}m_B$ (where $B$ stands now for the average wavelength $\langle\lambda\rangle=1736$\AA).
We computed magnification maps for each one of the images of the 27 pairs individually taking the $\kappa$ and $\gamma$
values from MED09 (except for  SBS0909+532 for which we take the values in Mediavilla et al. 2011a). 
We again consider only $\alpha=0.05$ and $\alpha=0.1$. The calculated maps are
$2000{\times}2000$ pixels in size, with a pixel size of 0.5 light-days. 
We have used the same logarithmic grid in $r_s$ from \S2,
as well as a linear grid $r^i_s=1.0+2.0{\times}i$ light days for $i=0\cdots 24$. 
Using the magnification maps, the probability of observing a
microlensing magnification $\Delta m^{obs}_j$ in image pair $j$ for a model with parameter $r_s^i$ is given by:
\begin{equation}
p_{r_s^i}({\Delta}m^{obs}_j){\propto}{\int}N_ie^{-\frac{1}{2}\chi^2}d{\Delta}m
\end{equation}
where
\begin{equation}
\chi^2=\frac{({\Delta}m-{\Delta}m^{obs}_j)^2}{\sigma^2_{{\Delta}m^{obs}_j}}.
\end{equation}

We can 
calculate a likelihood
function for $r_s$ as
\begin{equation}
L(r_s^i)\propto\prod_jp_{r^i_s}({\Delta}m^{obs}_j), 
\end{equation}
where $j$ runs over the 27 image-pairs considered.
\begin{figure}
\epsscale{1.1}
\plotone{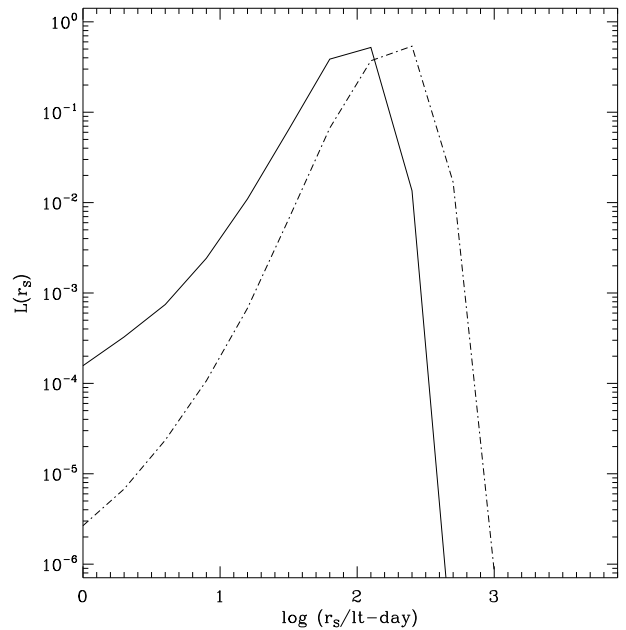}
\caption{Likelihood functions for the size of the accretion disk ${\log}\,r_s$ for $\alpha=0.05$ (continuous line) and $\alpha=0.1$ (dashed line) in the full sample of quasars by MED09.\label{fig2}}
\end{figure}

The results of this procedure for $\alpha=0.05$ and $\alpha=0.1$ are shown in 
Figure \ref{fig2}. The maximum likelihood occurs for ${\log}\,r_s=2.0^{+0.4}_{-0.8}$ 
and ${\log}\,r_s=2.3^{+0.4}_{-0.8}$ ($r_s$ in light days) for $\alpha=0.05$ and $\alpha=0.1$ respectively.
Taking into account that $r_s\propto{\langle}M{\rangle}^{1/2}$ where ${\langle}M{\rangle}$ is the mean mass
of the microlenses, our
results for the mean size of the accretion disk of quasars are $r_s=7.4^{+3.6}_{-4.1}({\langle}M{\rangle}/M_\odot)^{1/2}$ light days for $\alpha=0.05$ and $r_s=10^{+5.0}_{-5.5}({\langle}M{\rangle}/M_\odot)^{1/2}$
for $\alpha=0.1$. When scaled for a mean microlens mass of ${\langle}M{\rangle}=0.3M_\odot$, representative of the stellar populations in galaxies, the sizes of
the accretion disks become $r_s=4.0^{+2.0}_{-2.2}$ light days and $r_s=5.5^{+2.6}_{-3.1}$ light days for
 $\alpha=0.05$ and $\alpha=0.1$, respectively, at our average wavelength $\langle\lambda\rangle$=1736\AA. The intrinsic dispersion in $\lambda$ in between the sources implies an 
additional dispersion in $r_s$ of about 20\%. When taking this effect into 
account in the error bars, we finally have $r_s=4.0^{+2.4}_{-3.1}$ and $r_s=5.5^{+3.1}_{-3.7}$ light days for $\alpha=0.05$ and $\alpha=0.1$, respectively.

\begin{figure}[!t]
\epsscale{1.1}
\plotone{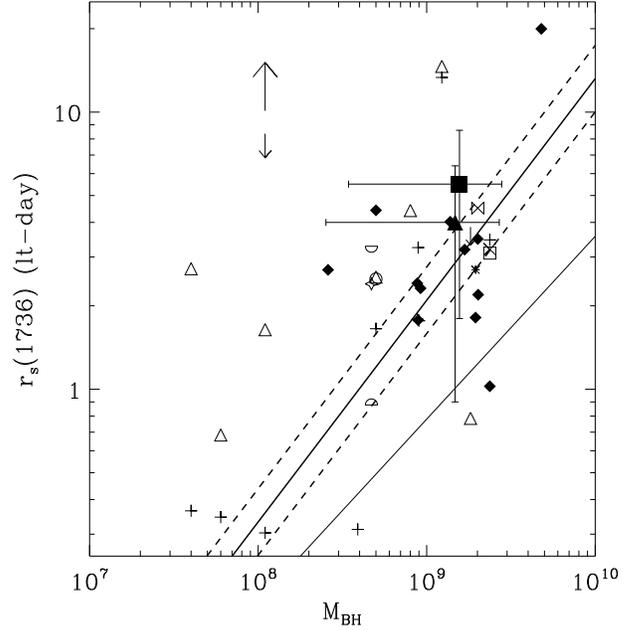}
\caption{Size of the accretion disk at 1736 \AA\ as a function of the mass of the black hole (from Mosquera \& Kochanek 2011) for $\langle M\rangle=0.3 M_\odot$ . 
Results from our joint analysis are the solid triangle and square for $\alpha$=0.05 and $\alpha$=0.1 respectively. Our estimates for individual objets are 
shown as filled diamonds (or upward arrows for lower limits).
Open triangles are from Blackburne et al. (2011a) 
for objects with
available black hole masses in Mosquera \& Kochanek (2011). The asterisk is the estimate for
SBS0909 from Mediavilla et al. (2011a). Estimates for HE1104 by Mu\~noz et al. (2011) and
Poindexter et al. (2008) are shown as an open square and an X respectively. The estimate for HE0435 by
Mosquera et al. (2011) is shown as an open circle. Upper half circle, lower half circle and star show the 
estimates for QSO2337 by Anguita et al. (2008), 
Kochanek (2004) and Eigenbrod et al. (2008) respectively. The estimate for Q0957 by Hainline et al. (2012) is shown as a bow tie. Downward arrows indicate the upper 
limits for MG0414 by Bate et al. 2008 (right) and
for SDSSJ0924 by Floyd et al. 2009 (left).
Thin line is the prediction from the thin disk model (with $L/L_E=1$ and $\eta$=0.1).
The thick solid line is the fit from Morgan et al. (2010) with their measurements shown
as pluses. The dashed
lines showing the limits due to the uncertainty in the y-intercept.\label{fig3}}
\end{figure}

Using equation (3) and a Bayesian approach with a lo\-ga\-rith\-mic prior on $r_s$ (see e.g.
Mediavilla et al. 2011a), we also made
individual estimates of $r_s$ for each one of the 
19 quasars in our sample. For those objects for which there are several image pairs, 
the probability distributions for the different pairs
were combined into a single probability function for that object. 
The resulting values for  ${\langle}M{\rangle}=0.3M_\odot$ and $\alpha=0.05$
are presented in Table \ref{tab1} and in Figure 3.
It is worth noting that the average value of these individual estimates 
(${\langle}r_s\rangle=4.2\pm4.4$ light days) is very close to the one obtained with the joint
likelihood analysis (although the use of a linear prior would increase the individual
size estimates by 80\%). Note that the average size shrinks noticeably (by ${\sim}30\%$) 
if the pairs with $|{\Delta}m_B|<0.1$ are excluded from the sample. 
Thus we confirm that favouring the selection of quasars with noticeable
 microlensing introduces a
clear bias towards smaller sizes (Kochanek 2004, Eigenbrod et al. 2008, Blackburne et al. 2011a). 
In Figure \ref{fig3} we compare our estimates with previous results.
Estimates from the present work are indicated with filled symbols, and
estimates from previous works with open symbols.
We have taken black-hole masses from Table 1 in Mosquera \& Kochanek (2011), which
they take from Peng et al. (2006), Greene et al. (2010) and Assef et al. (2011).
The thin line corresponds to the theoretical prediction for the thin disk model
at 1736 \AA\ (with $L/L_E=1$ and $\eta=0.1$).
The thick line corresponds to the empirical fit by Morgan et al. (2010) scaled to
$\langle\lambda\rangle$=1736 \AA\ using $p=4/3$
\begin{deluxetable}{llll}[t]
\tablecaption{Sizes for individual objects at 1736 \AA\label{tab1}}
\tablehead{
\colhead{Object} & \colhead{$r_s(1736)$\AA} & \colhead{$M_{BH}(10^9 M_\odot)$} & \colhead{Line}
}
\startdata
HE0047$-$1756 & $4.0^{+8.5}_{-2.7}$ & 1.38 & (CIV)\\ 
HE0435$-$1223 & $4.4^{+5.6}_{-2.4}$ & 0.50 & (CIV)\\ 
HE0512-3329 & $2.6^{+3.2}_{-1.4}$ & .... & \\
SDSS0806+2006& $1.4^{+1.9}_{-0.8}$& .... & \\
SBS0909+532 & $1.8^{+1.9}_{-1.0}$ & 1.95 & (H$\beta$)\\ 
SDSSJ0924+0219&$10.1_{-4.6}$\tablenotemark{a}& 0.11 &(MgII)\\
FBQ0951+2635& $1.8^{+2.9}_{-1.2}$ & 0.89 &(MgII)\\ 
QSO0957+561 & $3.5^{+7.3}_{-2.3}$& 2.01 &(CIV)\\ 
SDSSJ1001+5027&$4.6^{+10.5}_{-3.2}$ & .... & \\ 
SDSSJ1004+4112&$2.2^{+2.2}_{-1.1}$& 2.02 & (CIV) \\
QSO1017$-$207 &$ 3.2^{+6.02}_{-2.12}$ & 1.68& (CIV) \\ 
HE1104$-$1805 &$ 1.0^{+0.9}_{-0.5}$ & 2.37 &(CIV)\\ 
PG1115+080  &$ 2.3^{+2.6}_{-1.2}$& 0.92 & (CIV)\\
SDSSJ1206+432& $1.2^{+1.4}_{-0.6}$ & ....& \\ 
SDSSJ1353+1138&$7.5_{-4.4}$\tablenotemark{a} & ....& \\
HE1413+117  &$ 2.7^{+2.7}_{-1.3}$ & 0.26 &(CIV)\\ 
BJ1422+231  &$ 20.0^{17.3}_{-9.2} $& 4.79 & (CIV)\\
SBS1520+530 &$2.4^{+3.5}_{-1.5}$ & 0.88 &(CIV)\\ 
WFIJ0233$-$4723&$3.6^{+5.3}_{-2.2}$ & .... & \\ 
\enddata
\tablecomments{Sizes in light days for individual objects at 1736 \AA\  obtained from microlensing
magnifications of image pairs. Black hole masses are taken from the summary in Mosquera \& 
Kochanek (2011). Emission line used for the mass estimate is indicated in fourth column.}
\tablenotetext{a}{These values have been calculated for ${\Delta}m_B=0$ and are lower
limits rather than true estimates}
\end{deluxetable}

Our estimates are in good agreement with previous results (Blackburne et al. 2011a,
Mediavilla et al 2011a, Mosquera et al. 2011, Mu{\~n}oz et al. 2011, Morgan et al. 2010, Floyd et al. 2009, Anguita et al. 2008, 
Poindexter et al. 2008, Eigenbrod et al. 2008, Bate et al. 2008, and Kochanek 2004) and show a 
discrepancy with respect to the
predictions of the thin disk model of approximately a factor of 5.

\section{Conclusions}

We performed a statistical analysis to estimate the average size of the accretion
disks of lensed quasars from the large sample (MED09) of microlensing magnification measurements.
We find that most 
information on the size of the accretion disk is contained in the amplitude of the microlensing magnification and fairly independent of the amount of chromaticity. 

From a statistical analysis using measured microlensing magnification strengths 
from 27 image pairs from 19 lensed quasars in the sample of MED09
we measured the average size of the accretion disk of lensed quasars  at a rest wavelength of $\langle\lambda\rangle$=1736\AA. 
For a typical mean mass in the stellar population of the lens
of ${\langle}M\rangle=0.3M_{\odot}$, and taking into account the intrinsic dispersion in $\lambda$, 
we find 
 $r_s=4.0^{+2.4}_{-3.1}$ and  $r_s=5.5^{+3.1}_{-3.7}$ light days for
 $\alpha=0.05$ and $\alpha=0.1$ respectively. These estimates are in agreement with
other studies (Blackburne et al. 2011a,
Mediavilla et al 2011a, Mosquera et al. 2011, Mu{\~n}oz et al. 2011, Morgan et al. 2010, Floyd et al. 2009, Anguita et al. 2008, 
Poindexter et al. 2008, Eigenbrod et al. 2008, Bate et al. 2008, and Kochanek 2004) and again find that disks are larger than predicted by the thin disk theory.
We have also estimated the sizes of the 19 individual objects in the 
sample to find that they are in very good
statistical agreement with the result from the maximum likelihood analysis.

\acknowledgments
This research was supported by the Spanish Mi\-nis\-te\-rio de Educaci\'{o}n y Ciencia with the grants C-CONSOLIDER AYA2007-67625-C02-02, AYA2007-67342-C03-01/03, AYA2010-21741-C03/02. J.J.V. is also supported by the Junta de Andaluc\'{\i}a through the FQM-108 project. J.A.M. is also supported by the Generalitat Valenciana with the
grant PROMETEO/2009/64. C.S.K. is supported by NSF grant AST-1009756.

\,

\end{document}